\documentclass[aps,superscriptaddress,twocolumn,nofootinbib]{revtex4}
\usepackage{amsmath,amsbsy,amssymb,graphics}
\usepackage{graphicx,epsfig}
\def\beq{\begin{equation}}
\def\eeq{\end{equation}}
\def\bea{\begin{eqnarray}}
\def\eea{\end{eqnarray}}
\def\nnb{\nonumber}

\newcommand{\wti}{\widetilde}
\newcommand{\gsim}{\lower.7ex\hbox{$\;\stackrel{\textstyle>}{\sim}\;$}}
\newcommand{\lsim}{\lower.7ex\hbox{$\;\stackrel{\textstyle<}{\sim}\;$}}

\begin{document}

\title{ \vspace{-3ex}{\hfill hep-th/0605134 }\\[2mm] 
Anomaly inflow mechanism using Wilson line }                      

\author{  Wei Liao }
\affiliation{
TRIUMF, 4004 Wesbrook Mall, Vancouver, BC V6T 2A3, Canada
}
\begin{abstract}
It is shown that the anomaly inflow mechanism can be implemented
using Wilson line in odd dimensional gauge theories. An action
of Wess-Zumino-Witten (WZW) type can be constructed using
Wilson line. The action is understood in the odd dimensional bulk
space-time rather than in the even dimensional boundary. This 
action is not gauge invariant. It gives anomalous gauge variations 
of the consistent form on boundary space-times. So it can be used
to cancel the quantum anomalies localized on boundary space-times.
This offers a new way to cancel the gauge anomaly and construct
anomaly-free gauge theory in odd dimensional space-time.

\end{abstract}
\maketitle

When embedding an even ($2n$) dimensional space-time ${\cal M}$ 
into an odd ($2n+1$) dimensional space-time $\Sigma$ as a 
boundary, ${\cal M}=\partial \Sigma$, massless chiral fermion
can be localized on the even dimensional space-time ${\cal M}$.
Chiral fermion on ${\cal M}$ when coupled to gauge field can
induce gauge anomaly via radiative corrections and can give
potential problem to the gauge invariance.
The gauge invariance can be maintained and
the quantum anomaly on boundary can be canceled by anomalous gauge variation
of a classical action in the odd dimensional bulk space-time $ \Sigma$.
This is the anomaly inflow mechanism. The mechanism states that the 
anomalous gauge variation given by the bulk action flows into boundary and
cancels the anomalous gauge variation localized on the boundary.
Using this mechanism one can
construct consistent models in odd dimensional space-times
with localized chiral degrees of freedom on even dimensional 
boundaries. The conventional way to achieve the
cancellation is to use the Chern-Simons action in the bulk
~\cite{CH}. 

An important application of this mechanism is in condensed matter physics.
It is well known that one can understand the quantum Hall effect
using this mechansim ~\cite{wen,Bil}.
The mechanism has been generalized to more complicated
theories, {\it e.g.} the theory of gravitation and the
theory with gauge field of more than one indices.
They have been applied in string theory and the building
of extra dimensional models ~\cite{Bil}. 
In this article we show that there is a new way to
implement the anomaly inflow mechanism in odd dimensional
gauge theories, hence a new way to build anomaly-free gauge
theories in odd dimensional space-times. This is achieved 
using an action of WZW type constructed using Wilson line.
In the following we give a simple example to illustrate the
point of this article. We first introduce notations. 
Then we describe the anomaly
induced by chiral fermion localized on boundary.
The crucial bulk action to cancel anomaly is then introduced 
and its property is discussed.

\begin{figure}
\includegraphics[height=6cm,width=8cm]{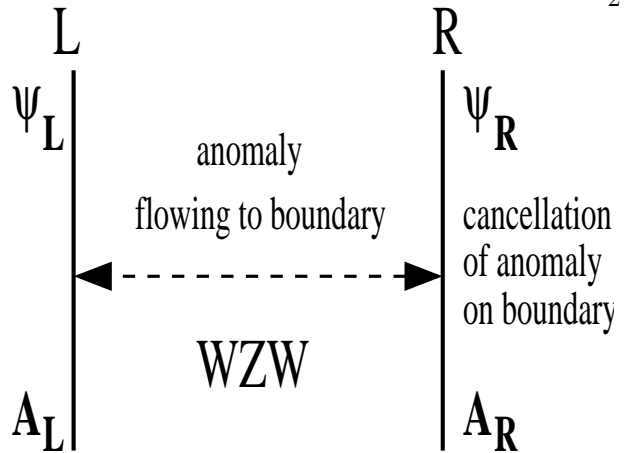}
\vskip 0.5cm
\caption{\small Anomaly inflow mechanism implemented by WZW action.
Gauge fields on L and R branes are induced by gauge field in the bulk:
$A_{L\mu}(x^\mu)=A_\mu(x^\mu,y=0)$ and $A_{R\mu}(x^\mu)=A_\mu(x^\mu,y=\pi R)$.}
\label{inflow}
\end{figure}

Consider an example on $\Sigma_3={\cal M}_2 
\times [0,\pi R]$ with coordinate $x^M=(x^\mu,x^2=y)$ ($\mu=0,1$).
${\cal M}_2$ is the two dimensional Minkowski space-time.
$\Sigma_3$ has two boundaries of type ${\cal M}_2$ at
$y=0$ and $y=\pi R$ separately. The gauge field $A_M=\sum_a T^a A^a_M$ of the
gauge group $G$ propagates in the bulk. $T^a$
is the generator of the group $G$. We define $A_M$ having
dimension [M]. Gauge coupling is absorbed into the gauge field.
We denote the boundary branes at $y=0$ and $y=\pi R$ as
brane L and brane R respectively. We introduce gauge fields on the
boundary branes, $A_{L\mu}$ and $A_{R\mu}$, which are
obtained when reducing $A_{\mu}$ to the boundaries:
\bea
A_{L\mu}(x^\mu)&&=A_\mu(x^\mu,y=0), 
\label{def1a} \\
A_{R\mu}(x^\mu)&&=A_\mu(x^\mu,y=\pi R).
\label{def1}
\eea
$A_{L,R\mu}$ live on two dimensional space-times 
and have no indices along $y$ direction.
Massless chiral fermions $\psi_L$ and $\psi_R$ 
are localized on brane L and brane R respectively.
They are minimally coupled to $A_L$ and $A_R$ separately,
see Fig. \ref{inflow}.

We study the property of gauge theory under
infinitesimal gauge transformation $U$ which approaches
unity at infinity:
\bea
U(x^\mu,y)=e^{i\epsilon},~~\epsilon(x^\mu,y)=\sum_a T^a \epsilon^a(x^\mu,y).
\label{def1b}
\eea
We have
\bea
\delta A_M= \partial_M \epsilon+ i[\epsilon, A_M].
\label{gaugt1}
\eea
$A_{L,R}$ transform as
\bea
\delta A_{L,R\mu}=\partial_\mu \epsilon_{L,R}
+i [\epsilon_{L,R},A_{L,R\mu}],
\label{gaugt2}
\eea
where $\epsilon_{L,R}$ are 
\bea
\epsilon_L(x^\mu)=\epsilon(x^\mu,y=0),~~
\epsilon_R(x^\mu)=\epsilon(x^\mu,y=\pi R).
\label{def2}
\eea

The 2D chiral gauge theory is anomalous
~\cite{jackiw,BZ}, that is the classical gauge symmetry
is broken at the quantum level.
The anomaly can be stated that the Noether current
derived from the classical Lagrangian is not conserved
when including radiative corrections of chiral fermions~\cite{jackiw,BZ}.
As an alternative it can be stated that the quantum action 
is not invariant under the gauge transformation 
of the classical Lagrangian ~\cite{Bil}.
In this 3D example it means the quantum
actions localized on two boundaries L and R are not invariant
under the gauge transformation given in (\ref{gaugt2}). 

We denote $\Gamma_L$ and $\Gamma_R$ as the quantum actions ~\footnote{
To avoid complication of renormalization, one needs
to limit the discussion to quantum action of one-loop level.}
\bea
\Gamma_L=\Gamma_{eff}\bigg|_{y=0}, ~~
\Gamma_R=\Gamma_{eff}\bigg|_{y=\pi R}.
\eea
Under the gauge transformation given in (\ref{gaugt2})
we have~\cite{Bil}
\bea
\delta\Gamma_L= -\frac{1}{4\pi} \int_{{\cal M}_2}
d^2x ~\varepsilon^{\mu\nu} ~Tr[\epsilon_L~ \partial_\mu A_{L\nu}],
\label{gaugt3} \\
\delta\Gamma_R= \frac{1}{4\pi} \int_{{\cal M}_2}
d^2x ~\varepsilon^{\mu\nu} ~Tr[\epsilon_R~ \partial_\mu A_{R\nu}],
\label{gaugt4}
\eea
where $\varepsilon^{01}=1$. (\ref{gaugt3}) and (\ref{gaugt4}) are
anomalous gauge variations on boundary space-times and are
of the form of consistent anomaly in 2D space-time ~\cite{BZ}.
$\pm$ signs arise from different chiralities in 2D theory~\cite{Bil}.

In a consistent gauge theory anomalies in (\ref{gaugt3}) and (\ref{gaugt4})
have to be canceled by contribution of other sector of the theory.
The cancellation of anomaly can be achieved when
an action in 3D space-time gives anomalies flowing
into the boundary space-times. The total anomaly can be arranged
to cancel. This is conventionally achieved using the Chern-Simons
action ~\cite{Bil}. We point out that we can achieve
anomaly cancellation using another action in the bulk space-time.
This bulk action is constructed using Wilson line and is
of WZW type.
Under the gauge transformation (\ref{gaugt2}) it gives
\bea
\delta \Gamma_{WZW}&&=\frac{1}{4\pi}\int_{{\cal M}_2}
\varepsilon^{\mu\nu} ~Tr[\epsilon_L ~\partial_\mu A_{L\nu}] \nnb \\
&& -\frac{1}{4\pi}\int_{{\cal M}_2}
\varepsilon^{\mu\nu} ~Tr[\epsilon_R ~\partial_\mu A_{R\nu}].
\label{gaugt7}
\eea
It gives the consistent form of anomaly on two
boundaries. It is clear that the gauge invariance 
is maintained in the action
\bea
\Gamma=\Gamma_{WZW}+\Gamma_L+\Gamma_R.
\label{action2}
\eea
The construction of the theory is shown in Fig. \ref{inflow}.
We emphasize that this is a new way to
achieve anomaly cancellation in 3D gauge theories.
In the following we elaborate on this WZW type action
in 3D space-time.

Note that the original WZW action is constructed in non-linear
sigma model and is an action in 4D space-time 
~\cite{wznw,cgws,krs,kt,mm,manes}. Extension of this
action to even dimensional space-time exists in the literature~\cite{wu}.
$\Gamma_{WZW}$ used in this article should be
understood living in three (odd) dimensional space-time, hence is
a further extension of the original WZW action. This extension is
achieved using the Wilson line along the third dimension
~\cite{liao,Hill,hill}.
We introduce Wilson line
\bea
W_y(x^\mu,y)= {\cal P} e^{i\int^y_0 dy' ~A_2(x^\mu,y')},
\label{WilL1}
\eea
and 
\bea
W(x^\mu)= {\cal P} e^{i\int^{\pi R}_0 dy' ~A_2(x^\mu,y')}.
\label{WilL2}
\eea
${\cal P}$ is the path-ordering operator.
$A_2$ is the gauge field along the third dimension.
Since $W_y(x^\mu,y=0)=1$ the configuration of $W_y$ is
the mapping of ${\wti \Sigma}_3$ to the space of the gauge group $G$.
${\wti \Sigma}_3$ is $\Sigma_3$ with the 2D boundary
at $y=0$ shrinking to a point. So ${\wti \Sigma}_3$
has a single boundary: $\partial {\wti \Sigma}_3={\cal M}_2$.

$W=W_y(y=\pi R)$ transforms as a bifundamental of gauge
transformations on two boundary space-times:
\bea
W'(x^\mu)=U(x^\mu,y=0) W(x^\mu) U^{-1}(x^\mu,y=\pi R).
\label{gaugt6a}
\eea
$W$ is a link field which can mix the degrees of the freedom
on two boundaries. Using it we can write an action~\cite{liao}
\bea
\Gamma_{WZW}&&=-\frac{1}{12\pi} \int_{\Sigma_3} d^3x ~\varepsilon^{RST}
~ Tr[(\partial_R W_y) W^{-1}_y \nnb \\
&& \times (\partial_S W_y) W^{-1}_y
(\partial_T W_y) W^{-1}_y ] \nnb \\
&&-\frac{i}{4\pi}\int_{{\cal M}_2} d^2x ~ \varepsilon^{\mu\nu}
 ~Tr[A_{L\mu}W_{L\nu} \nnb \\
&&+A_{R\mu}W_{R\nu}
-i A_{R\mu}W^{-1} A_{L\nu} W],
\label{action}
\eea
where\footnote{There is a difference on sign compared to
the action written in Ref. \cite{liao}.} 
$R,S,T$ run over $0,1,2$, $\varepsilon^{012}=1$ and
\bea
W_{L\mu}= (\partial_\mu W)W^{-1}, ~~W_{R\mu}=W^{-1}(\partial_\mu W).
\label{def4}
\eea

The action written in
(\ref{action}) is formally of WZW type and
is similar to the WZW action in 2D space-time.
This action has to be interpreted as living in 3D space-time,
quite different from the interpretation of the usual WZW action
living in even dimensional space-time~\cite{wznw}.
One can understand this interpretation by noting that
this action gives non-local interaction for gauge fields $A_L$
and $A_R$ on different boundary branes. That is it gives
interaction for degrees of freedom at space-like distance.
\footnote{This non-locality along the compact space-like
dimension may not affect the 2D locality. 
A proof for this statement is needed.} 
In particular, the last term in (\ref{action}) can not
be understood as living in any one of the two boundary branes.
We emphasize that this action can not be understood as
localized on any one of the boundary branes.

Gauge transformation property of (\ref{action}) can be easily 
computed. 
Note that under the infinitesimal transformation (\ref{def1b})
and (\ref{gaugt1}) we have
\bea
\delta W_y(x^\mu,y)=i\epsilon_L W_y(x^\mu,y) - i W_y(x^\mu,y) \epsilon_y,
\label{gaugt6}
\eea
where
\bea
\epsilon_y(x^\mu)=\epsilon(x^\mu,y).
\label{def3}
\eea
Using (\ref{gaugt2}) and (\ref{gaugt6}) we can easily get
the transformation property of $\Gamma_{WZW}$. 
Dropping the surface integration
at infinity  Eq. (\ref{gaugt7}) is obtained.

The above construction in 3D can be generalized to
5D space-times with boundaries. One can also take the
picture shown in Fig. \ref{inflow}. $\psi_L$ and
$\psi_R$ are then understood as localized on 4D
boundaries. Similarly we have quantum actions
$\Gamma_L$ and $\Gamma_R$ on two boundary branes
which are obtained including
radiative corrections of $\psi_L$ and $\psi_R$ separately.
Gauge anomaly in 4D space-time states
that $\Gamma_L$ and $\Gamma_R$ are not gauge
invariant~\cite{Bil,liao}. For an infinitesimal gauge transformation, they
give
\bea
\delta \Gamma_L&&= -\frac{1}{24\pi^2} \int_{{\cal M}_4} d^4x
~\omega^1_4(A_{L\mu},\epsilon_L),
\label{gaugt8} \\
\delta \Gamma_R&&= \frac{1}{24\pi^2} \int_{{\cal M}_4} d^4x
~\omega^1_4(A_{R\mu},\epsilon_R),
\label{gaugt8a}
\eea
where 
\bea
\omega^1_4(B_\mu,\varepsilon)= \varepsilon^{\mu\nu\rho\sigma}
Tr[\varepsilon ~\partial_\mu (B_{\nu}\partial_\rho B_{\sigma}
-\frac{i}{2}B_{\nu}B_{\rho}B_{\sigma}) ].
\label{gaugt8b}
\eea
$\mu,\nu,\rho,\sigma$ run over $0,1,2,3$ and
$\varepsilon^{0123}=1$. (\ref{gaugt8}) and (\ref{gaugt8a})
are of the form of consistent anomaly in 4D space-time~\cite{BZ}.

A 5D WZW type action has been constructed using Wilson line in 
Ref. \cite{liao,hill}. It gives
anomalous gauge variations flowing into the boundary branes.
Readers are referred to Ref. \cite{liao}
for explicit formula (an extra minus sign is needed
to get the gauge variation (\ref{gaugt9}) in the following).
The action constructed is similar to the 4D
WZW action~\cite{wznw,cgws,krs,kt,mm,manes}. However it has to be
understood in the 5D bulk rather than localized on any one of
the boundary space-times. This action gives non-local
interaction for gauge fields $A_L$ and $A_R$ of two
boundary space-times. Under the infinitesimal gauge transformation
it gives
\bea
\delta \Gamma_{WZW}&&=\frac{1}{24\pi^2} \int_{{\cal M}_4} d^4x
~\omega^1_4(A_{L\mu},\epsilon_L) \nnb \\
&&-\frac{1}{24\pi^2} \int_{{\cal M}_4} d^4x
~\omega^1_4(A_{R\mu},\epsilon_R)
\label{gaugt9}
\eea
The gauge variation on boundaries given by this
action is of the form of consistent anomaly.
It is clear that anomalous gauge variations of $\Gamma_L$ and $\Gamma_R$
are canceled by the gauge variation of the bulk action.
This implements the anomaly inflow mechanism using
Wilson line in 5D space-time.

This is the story to implement the anomaly inflow mechanism 
using Wilson line in 3D and 5D gauge theories.
Using this mechanism one can construct anomaly-free gauge theories 
in 3D or 5D space-time with massless chiral fermions
localized on boundaries, as shown in Fig. \ref{inflow}.
We emphasize that this new mechanism is quite different
from the old one using the Chern-Simons action. Because of
the non-local nature of $\Gamma_{WZW}$ we do not expect to
find a current along the compact space-like dimension. 
On the contrary, the old mechanism indeed has a current 
along the compact dimension.
This current is derivable from the Chern-Simons term.

As a further comparison we note that the WZW-type term 
used in this article and the Chern-Simons term have
different origins in effective theory.
In a previous work \cite{liao}
we have shown how this WZW type action can be completely derived when
integrating out some chiral fermions on even dimensional boundaries. 
These chiral fermions are localized on boundary space-times
and are coupled to Wilson line via non-local Yukawa-type
interaction. Integrating out these chiral fermions we find
the WZW type action $\Gamma_{WZW}$ appearing in the effective action.
On the other hand, it is known for a long time that the Chern-Simons
term appears in effective theory when integrating out
fermion in odd dimensional space-time~\cite{redlich}.
It is clear that $\Gamma_{WZW}$ used in this article is
different from and independent of the Chern-Simons
action. This opinion is different from the opinion taken in ~\cite{Hill}.
We emphasize that the mechanism shown in this article is a new mechanism,
not the old mechanism rewritten in a new form.

In summary we have shown that there is a new type of anomaly inflow
mechanism implemented 
by a WZW type action constructed using Wilson line.
This action has to be understood as living in odd dimensional space-time
rather than in even dimensional space-time.
This point is also emphasized in Ref. \cite{liao}. 
The action constructed is not gauge invariant and it gives
anomalous gauge variations of the consistent form on
boundary space-times. These anomalous gauge variations
cancel the anomalous gauge variations localized on
boundaries. Using this mechanism the gauge invariance of the odd
dimensional gauge theory can be achieved when massless
chiral fermions are localized on boundary space-times.

We expect that the 3D version of this new mechanism is of potential
interests to condensed matter physics. It may have application
similar to the application
of the old mechanism using Chern-Simons term in quantum
Hall effects. However the experimental devices need to
be different. This topic requires further research.
We expect that this mechanism can be generalized
to odd dimensional space-times of dimension larger than five.
Action to implement the mechanism will be similar to
the WZW action in even dimensional
space-times of dimension larger than four ~\cite{wu}.
Examples shown in this article are for gauge theory in flat
space-time. We expect this mechanism can be generalized
to curved background space-time. Using this mechanism
we are able to construct new type of anomaly-free gauge theories
in odd dimensional space-times. Extension of this
mechanism to more complicated theories, {\it e.g.}
the theory of gravity, requires further research.
\\

Acknowledgment: the author wishes to thank C. S. Lam
for helpful comments.

\end{document}